\documentclass[aps,prl,twocolumn,amsmath,amssymb,superscriptaddress,reprint]{revtex4-1}
\usepackage{enumerate}
\usepackage{amsmath}
\usepackage{graphicx}
\usepackage{color}
\usepackage[breaklinks]{hyperref}
\usepackage{psfrag}
\usepackage{stmaryrd}
\usepackage{amssymb}
\usepackage[latin1]{inputenc}
\usepackage[english]{babel}
\usepackage{float}
\def \degree {^\mathrm{o}}
\let\oldtabular=\tabular
\def\tabular{\small\oldtabular}
\usepackage{mathrsfs}
\usepackage[margin=3cm]{geometry}
\usepackage{array}

\def \degree {^\mathrm{o}}

\begin{document}

\author{Marie-Julie Dalbe}
\affiliation{Laboratoire de Physique de l'\'{E}cole Normale
Sup\'{e}rieure de Lyon, CNRS and Universit\'{e} de Lyon, 69364 Lyon, France}
\affiliation{Institut Lumi\`ere Mati\`ere, UMR5306 Universit\'e
Lyon 1-CNRS, Universit\'e de Lyon, 69622 Villeurbanne, France}

\author{Pierre-Philippe Cortet}
\affiliation{ Laboratoire FAST, CNRS, Univ Paris Sud, Universit\'e  Paris-Saclay, 91405 Orsay, France}

\author{Matteo Ciccotti}
\affiliation{Laboratoire SIMM, UMR7615 ESPCI-CNRS-UPMC-PSL, 75231 Paris, France}

\author{Lo\"{i}c Vanel}
\affiliation{Institut Lumi\`ere Mati\`ere, UMR5306 Universit\'e
Lyon 1-CNRS, Universit\'e de Lyon, 69622 Villeurbanne, France}

\author{St\'{e}phane Santucci}
\email{stephane.santucci@ens-lyon.fr}
\affiliation{Laboratoire de Physique de l'\'{E}cole Normale
Sup\'{e}rieure de Lyon, CNRS and Universit\'{e} de Lyon, 69364 Lyon, France}

%\date{\today}
\title{Multi-scale stick-slip dynamics of adhesive tape peeling}

\begin{abstract}
Using a high-speed camera, we follow the propagation of the
detachment front during the peeling of an adhesive tape from a flat surface. 
In a given range of peeling velocity, this
front displays a multi-scale unstable dynamics, entangling two
well-separated spatio-temporal scales,  
which correspond to microscopic and macroscopic dynamical stick-slip instabilities.
While the periodic release of the stretch energy of the whole peeled
ribbon drives the classical macro-stick-slip, we show that the micro-stick-slip, 
due to the  regular propagation of transverse dynamic fractures
discovered by Thoroddsen \textit{et al.} [Phys. Rev. E \textbf{82}, 046107 (2010)], 
 is related to a high-frequency periodic release of the elastic bending energy of the adhesive ribbon concentrated  
 in the vicinity of the peeling front. 
\end{abstract}

%\pacs{62.20.mm,68.35.Np,82.35.Gh}
%62.20.mm Fracture
%68.35.Np Adhesion
%82.35.Gh Polymers on surfaces; adhesion

\maketitle

Understanding the velocity at which fractures propagate is a key issue 
in various fields, ranging from mechanical engineering to seismology.
When fracture energy becomes a decreasing function of crack velocity ({\it i.e.}
it costs less energy for a crack to grow faster), a dynamical instability 
can occur, leading to crack velocity periodic oscillations, 
as for example in friction problems~\cite{Baumberger1994}.
During these oscillations, 
transitions from quasi-static to dynamic 
crack propagation regimes 
may then develop~\cite{Svetlizky2014}, the latter being prone to trigger fracture front instabilities~\cite{Fineberg1999}.
The stick-slip oscillations of the detachment front
during the peeling of adhesive tape is another outstanding example, extensively  studied~\cite{Gardon1963,Aubrey1969,Racich1975,Barquins1986,Maugis1988,Ryschenkow1996,Gandur1997,Barquins1997,Ciccotti2004,De2004,Cortet2007,De2008}.    
However, it remains nowadays an industrial concern, leading
to unacceptable noise levels,
damage to the adhesive and mechanical problems on assembly
lines. Thanks to technological progress in high-speed imaging, 
direct observations of the peeling front
dynamics in the stick-slip regime revealed 
a complex dynamical
and multi-scale process~\cite{Thoroddsen2010,Cortet2013,Marston2014,Dalbe2014,Dalbe2014b}, 
challenging our current understanding. 
In particular, 
Thoroddsen \textit{et al.}~\cite{Thoroddsen2010} observed a regular substructure of few hundred microns wide transverse bands, formed during the ``slip" phase of the stick-slip oscillations by a periodic propagation of rapid fractures
across the tape width, 
intrinsically akin to a  dynamic fracture instability~\cite{Fineberg1999}.

In this Letter, we unveil the physical mechanism of 
this
instability.
We show that the micro-stick-slip dynamics of the detachment front, 
due to the high-frequency periodic propagation of transverse fractures, 
can be observed (i) for
imposed peeling velocities in the macroscopic stick-slip domain,
during the slip phase of the macroscopic instability, but also, (ii) for imposed peeling velocities in a
finite range beyond the macroscopic stick-slip domain, where the
peeling is rapid and regular at the macroscopic scale. 
In the former case, we confirm here the entanglement of the microscopic
and macroscopic dynamical instabilities leading to a complex
multi-scale stick-slip dynamics~\cite{Thoroddsen2010,Marston2014}.
In contrast with the macroscopic stick-slip, we find that the high-frequency micro-stick-slip  has an amplitude and a period independent of the peeled tape length $L$ between the detachment front and the point at which the traction is applied.
Thanks to time-resolved measurements of the peeled tape Elastica profiles, we show that 
the microscopic instability is driven by a periodic release of the ribbon bending energy concentrated in the vicinity of the peeling front. 
From the measurements of the bending energy release during a micro-slip, we could 
estimate the effective peeling fracture energy in the micro-stick-slip regime, in agreement with direct macroscopic measurements~\cite{Barquins1997}. 
\begin{figure}[h!]
\centerline{\includegraphics[width=6.25cm]{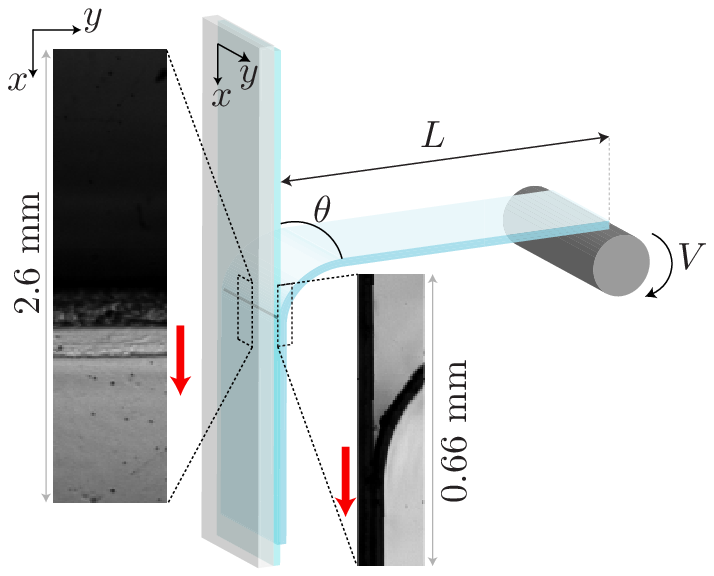}}
\vspace{-0.35cm}
\caption{Microscopic observations of the detachment front and
ribbon profile when peeling a tape from a flat surface
at a velocity $V$, a distance $L$ and an angle $\theta \simeq 90 \degree$. The arrows show the propagation direction.}\label{fig:schema_exp}
\end{figure}

We peel an adhesive tape from a microscope plate by winding its extremity 
at a constant velocity $V$ using a brushless motor.
The winding cylinder is at a distance $L$ from the substrate and
the peeling angle is set to $\theta \simeq 90 \degree$ (Fig.~\ref{fig:schema_exp}). 
A high-speed camera (PHOTRON SA5) mounted
on a microscope images 
the substrate-adhesive interface
over about 1~mm$^2$ at a fixed spatial resolution of
$9.5$~$\mu$m/px. 
Depending on the imposed peeling velocity $V$,
frame rate is varied between $300\,000$~fps for
$256\times 64$~px$^2$ images and $525\,000$ fps for $192\times
32$~px$^2$ images. During an experiment, the peeling angle
$\theta$ and length $L$ vary less than $4\%$ and
$0.2\%$ respectively, and are thus considered constant. We use
3M~Scotch$^{\circledR}$~600~\cite{Barquins1997,Cortet2007,Cortet2013,
Dalbe2014,Dalbe2014b} made of a polyolefin blend backing
($e=34~\mu$m thick, $b=19$~mm wide, tensile modulus $E = 1.41\pm
0.11$ GPa) coated with a 20 $\mu$m layer of a synthetic acrylic
adhesive, 
peeled from a plate
previously covered by one layer of Scotch$^{\circledR}$~600 tape
with its backing release coating cleansed with ethanol.

\begin{figure}[h!]
\centerline{\includegraphics[width=6.25cm]{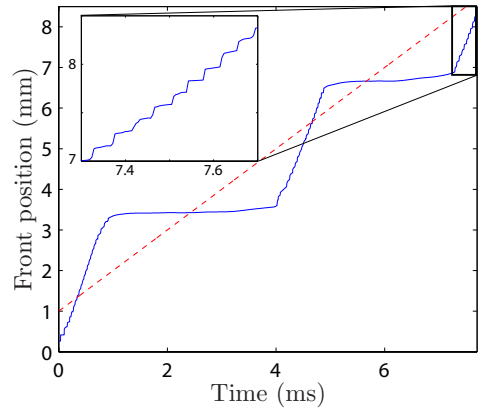}}
\vspace{-0.4cm}
\caption{Local peeling front position (in direction $x$) as a function of time for
$L=80$~cm and $V=1$~m~s$^{-1}$, showing  two macro-stick-slips. A zoom on a slip event reveals the micro-stick-slip instability.
 The slope of the dashed
line is the driving velocity $V$.}\label{fig:pos}
\end{figure}

Analyzing the grey levels of the recorded images (Fig.~\ref{fig:schema_exp}, left),  
we extract the longitudinal position of the detachment front $x(y_0,t)$ at a given transverse position $y_0$. 
Figure~\ref{fig:pos} gives a typical example of such local front position time series~\cite{video}. 
While the average front velocity over the experiment duration is equal to the driving velocity 
$V$ (dashed line in Fig.~\ref{fig:pos}), we observe 
that the peeling front advances by steps,
characteristics of the stick-slip instability. 
We could observe this
standard 
macroscopic instability with periods of the front velocity oscillations of a few to a few tens of milliseconds
over a finite range of driving
velocity $V \in [V_a,V_d]$~\cite{Dalbe2014b}.
In this range, the limit cycles of the macroscopic instability progressively change from typical stick-slip relaxations to nearly sinusoidal oscillations,
controlled by the peeled tape inertia~\cite{Dalbe2014b}. 
The instability onset observed here at 
$V_{a}=0.35\pm 0.05$~m~s$^{-1}$ 
corresponds to the velocity at which 
the fracture energy $\Gamma$ of the adhesive-substrate joint
starts to decrease with $V$~\cite{V_a}.
For $V>V_{d}=5.6\pm1.4$~m~s$^{-1}$, 
we observe anew a stable peeling 
at the
millisecond time-scale~\cite{Dalbe2014b}.
During the rapid slip phases, we observe  a
secondary regular step-like dynamics at much shorter spatial and temporal
scales, as shown in Fig.~\ref{fig:pos}. 
Pushing the acquisition rate up to $700\,000$~fps, 
we observe~\cite{video} that this microscopic stick-slip dynamics of the local peeling front position is due to the periodic propagation of fractures at the substrate-adhesive interface in the direction $y$ transverse to the peeling direction $x$.
The width (in direction $x$) of the adhesive bands detached from the substrate by the propagation of those transverse fractures sets the micro-stick-slips amplitude $A_{mss}$. 
Such regular propagation of rapid fractures during the slip phase of the peeling instability was discovered by Thoroddsen \textit{et al.}~\cite{Thoroddsen2010}.
In contrast to our experiments, they 
report tilted peeling fronts (up to $30\degree$) with respect to the transverse direction $y$,  
probably due to their manual pulling procedure.
We estimate similar 
transverse fracture velocities from $650$~m~s$^{-1}$ up to $900$~m~s$^{-1}$ confirming the dynamic nature of this rupture process. 

Contrary to the macro-instability, the driving velocity $V$ is not a relevant parameter to 
determine the presence of the micro-instability,
since  for 
$V \in [V_a,V_d]$,  the micro-stick-slip is observed during the macro-slips but not during the macro-sticks. 
A possible order parameter for this micro-instability is the peeled tape velocity beyond the tape bended section close to the detachment front. We estimate this velocity by measuring the detachment front velocity $v_m(t)=\int_{t}^{t+\tau}\dot{x}(y_0,t')dt'/\tau$ averaged at a time-scale $\tau$ of a few hundreds of $\mu$s, intermediate between the time-scales of the macro and micro stick-slip dynamics. The separation of those time-scales by at least 1.5 order of magnitude in our experiments ensures the robustness of 
the observable $v_m(t)$.
For $V \notin [V_a,V_d]$, the peeling is macroscopically stable 
at the ms time scale.
 At the $\mu$s time scale, a stable peeling is still observed, with $\dot{x}(t)\simeq v_m(t)=V$, when $V <V_a$ or $V>v_d=21.1 \pm 3.1$~m~s$^{-1}$.
 However, for $V \in [V_d, v_d]$, micro-stick-slip is observed, such that $\dot{x}(t)\neq v_m(t)$ and $v_m(t)=V$.
For $V \in [V_a,V_d]$, the front alternatively propagates either at a velocity  $v_m(t)$ smaller than $V_a$ (during the macro-stick phase) for which peeling is stable at the $\mu$s scale ($\dot{x}(t)\simeq v_m(t)$), 
or with a high-frequency jerky dynamics -- the micro-stick-slip -- of mean velocity $v_m(t)$ larger than $V_a$ (during the macro-slip phase, see inset of Fig.~\ref{fig:pos}).
As a result,  when $V_a <v_m(t)<v_d$, 
the front always displays micro-stick-slips,  independently of the macroscopic peeling regime (see the state diagram in supplementary material).
Thus, we report that the macro and micro instabilities appear for similar velocities,  
while they disappear at significantly different ones, $V_d$ and $v_d(>V_d)$ respectively.
Interestingly, the
characteristic velocity $v_d$ at which
micro-stick-slip disappears is close to the velocity $V_0 \sim 19$~m~s$^{-1}$~\cite{Barquins1997} 
above which the 
fracture energy of the adhesive-substrate joint $\Gamma$
is increasing anew with the front velocity $v_m$. 
Our results tend to show 
that, while the macro-stick-slip instability does not develop
over the whole decreasing range of $\Gamma$~\cite{Yamazaki2006,Dalbe2014b}, 
the micro-stick-slip instability seems to occur as soon as the velocity $v_m(t)$ of the detachment front belongs to this
decreasing branch of $\Gamma$, 
and thus,  is fundamentally a dynamic fracture instability.
\begin{figure}[h!]
\centerline{\includegraphics[width=6.25cm]{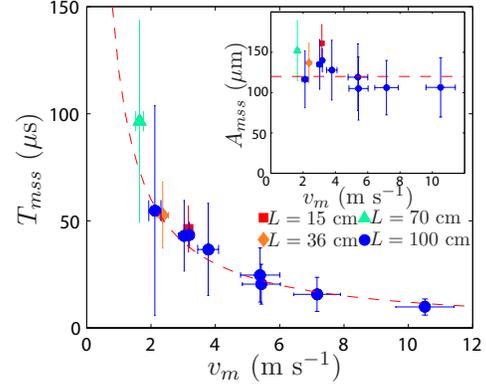}}
\vspace{-0.375cm}
\caption{Mean micro-stick-slip duration $T_{mss}$ and amplitude
$A_{mss}$ (inset) as function of $v_m$ 
for various $L$. 
The dashed lines represent
$T_{mss}=\langle A_{mss}\rangle/v_m$ and
$A_{mss}=\langle A_{mss}\rangle$ (inset), with $\langle
A_{mss}\rangle = 120~\mu$m. }\label{fig:ATvsV}
\end{figure}

For a set of parameters ($L,V$),  
we extract amplitude  and duration of
at least a hundred of
micro-stick-slips cycles from a
dozen of experiments. We compute the mean amplitude $A_{mss}$, mean duration $T_{mss}$,  
and corresponding mean front velocity $v_m$.
Contrary to the
macro-instability, where durations and amplitudes of the cycles are proportional to $L$~\cite{Maugis1988, Dalbe2014},  the
micro-stick-slips features appear not to depend significantly on $L$ (Fig.~3).
The instability amplitude $A_{mss}$ is moreover nearly constant with $v_m$  
with a mean
value  $\langle A_{mss}\rangle = 120\pm
1~\mu$m (inset of Fig.~\ref{fig:ATvsV}). 
Consequently, the 
micro-stick-slip period scales as 
$T_{mss}=\langle A_{mss}\rangle/v_m$ (dashed line in Fig.~\ref{fig:ATvsV})~\cite{Marston2014}.
Thoroddsen \textit{et al.}~\cite{Thoroddsen2010} found a larger constant value, probably due to the different adhesive-substrate joint studied.
We also note that, for $v_m>5\pm 1$~m~s$^{-1}$, 
the micro-stick-slip period 
can become smaller than the time for a transverse fracture to cross the tape width.
Thus, for these $v_m$, several transverse fractures shifted of $T_{mss}$ may
be observed~\cite{Thoroddsen2010}.

\begin{figure}[b!]
\centerline{\includegraphics[width=6.5cm]{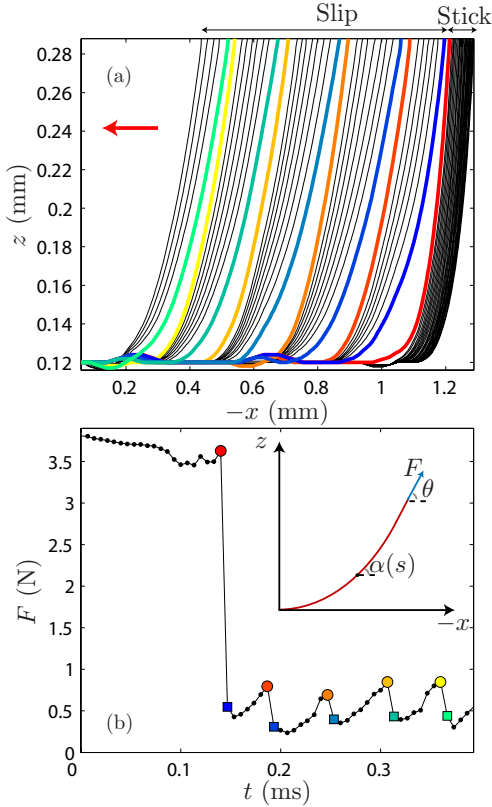}}
\vspace{-0.4cm}
\caption{
Ribbon profile extracted at intervals of $6.7~\mu$s (a) 
and corresponding peeling force time evolution (b)
when peeling at $V=0.73$~m/s,
$L=1$~m. }\label{fig:profils}
\end{figure}
We performed additional experiments in order to image 
the peeled tape profile (Fig.~\ref{fig:schema_exp}) at $150\,000$~fps and
$512\times80$~px$^2$, with a resolution of $5$~$\mu$m/px.
 Fig.~\ref{fig:profils}(a) shows the profiles extracted at  intervals of $6.7~\mu$s
 for an experiment at $V=0.73$~m/s and $L=1$~m. 
After the macro-stick phase during which the peeling front propagates slowly (about $0.04$~m~s$^{-1}$, here), leading to a dense region of profiles on the right of Fig.~\ref{fig:profils}(a), we observe
the periodic micro-stick-slip dynamics during the macro-slip phase:  
the tape curvature increases, before
decreasing suddenly in less than $6.7~\mu$s, and so on. 
This drop of the ribbon curvature implies a release of its bending energy. 
In order to estimate the corresponding local energy release, 
we compute for each profile
the angle $\alpha$ between the substrate plane and the tangent to
the peeled tape as a function 
of the curvilinear abscissa $s$ from the detachment front (as defined in the inset of Fig.~\ref{fig:profils}(b)). 
Using the Elastica theory~\cite{Love1944} (with a simple clamping boundary condition at the peeling front $s=0$),  for a 
force $F$ pulling the  tape extremity with an angle $\theta$ (see supplementary material), one can show that,
$\alpha(s) = \theta-4\arctan\left(e^{-s/r}\tan\frac{\theta}{4}\right)$,
where   
$r=\sqrt{EI / F}$,  with $E$ the ribbon tensile modulus, 
and $I=e^3b / 12$ its second area moment. 
Fitting each experimental profile $\alpha(s)$,
we extract the adjustable parameter, the peeling force $F$, as a function of time.
 A typical
example of its evolution is shown in Fig.~\ref{fig:profils}(b) for
the experiment of Fig.~\ref{fig:profils}(a).
At the end of the macro-stick,  the force suddenly drops from about $3.5$~N to approximately $0.5$~N
at  the beginning of the macro-slip phase. Then, the peeling
force $F(t)$ displays quasi-periodic oscillations with slow increases  
followed by sudden drops of
amplitude typically about $0.4$~N. 
These force oscillations are synchronized with the periodic propagation of the transverse fractures and therefore with the whole micro-stick-slip dynamics.
Their amplitude is  nearly independent of the front velocity $v_m$ (checked up to $4$~m~s$^{-1}$ only, due to 
difficulties to image properly the tape profiles at larger $v_m$). 
For each micro-stick-slip cycle, we compute the variation of the bending energy $\Delta E_c$ associated to the drop of the peeling force using $E_c=4\sqrt{E\,I\,F}\sin^2\theta/4$~\cite{Roman2013}. 
Assuming that this energy release allows the front to advance of a step $A_{mss}$, 
we estimate an effective fracture energy associated to the micro-stick-slip regime as $\Gamma_{mss}=\Delta
E_c/(b\,A_{mss})$. 
We measure  $\Gamma_{mss}=13.5\pm1.3$~J/m$^{2}$, independently of $v_m$, 
(neither $A_{mss}$ nor $\Delta E_c$ 
vary significantly with $v_m$),
which is close to the steady state macroscopic fracture energy $\Gamma=G(V)$
measured at the local minimum of $G(V)$  at $V \simeq 19$~m~s$^{-1}$~\cite{Barquins1997}.

Interestingly, at the scale of the tape curvature close to the detachment front $r = \sqrt{EI / F} \simeq 330\,\mu$m, just before the force drops, 
 the ribbon of linear mass density
$\mu=(8.5\pm0.15)\,10^{-4}\,\rm{kg~m}^{-1}$ 
has a natural bending vibration frequency of the order of 
$\sqrt{E I/\mu r^4} = 93\,$kHz~\cite{Weaver1990}. 
This characteristic 
 frequency sets a lower bound for the time-scale for the release of the tape bending energy,
 in agreement with the highest micro-stick-slip frequency we measured.  
We also note that, following the force drop at the end of the macro-stick, a 
decompression wave will propagate along the tape at $c=\sqrt{Eeb/\mu}\simeq 1000$~m~s$^{-1}$, converting elastic stretch energy of the tape into kinetic energy behind the wave front.
This kinetic energy allows for a local reloading of the ribbon at an almost constant velocity after each micro-slip.

To conclude, we have shown that the  small-scale high-frequency instability of adhesive tape peeling
corresponds to a micro-stick-slip dynamics fundamentally different from the macroscopic stick-slip instability.
The regular propagation of rapid transverse cracks which creates the micro-stick-slip spatio-temporal pattern 
appears as   a purely dynamic fracture instability.
We have shown that the velocity $v_m$ of the peeled tape just beyond the bending region close to the peeling front is the control parameter of the micro-stick-slip instability, 
which is driven by the periodic release of the bending energy.  
The bending elasticity allows for the detachment front velocity to be instantly different from the tape velocity $v_m$ beyond the bending region (but equal on average).
Further work is needed to 
understand the interplay between the release of the elastic stretch energy of the whole free standing tape, driving the macro-stick-slip, and the fast release of the ribbon bending energy close to the peeling front.

\acknowledgments

We thank D. Le Tourneau for his help in the design of the set-up, O. Ramos for his help in the early stage of the
experiments,  C. Creton and R. Villey for enlightening discussions.
This work has been supported by the Research  Federation ``A. M.
Amp\`ere" (FRAMA) and the French ANR through Grant
``STICKSLIP'' No. 12-BS09-014.


\begin{thebibliography}{5}


\bibitem{Baumberger1994}
T. Baumberger and C. Caroli, 
Advances in Physics, 55:3-4, 279-348 (2006).


\bibitem{Svetlizky2014}
I. Svetlizky and J. Fineberg, 
%Classical Shear Cracks Drive the Onset of Dry Frictional Motion, 
Nature, 509,  205-208 (2014)


\bibitem{Fineberg1999}
J. Fineberg and M. Marder, 
%"Instability in Dynamic Fracture",
  Physics Reports, 313, 1-108 (1999).

\bibitem{Gardon1963}
{J.L. Gardon}, {J. Appl. Polym. Sci.}, \textbf{7}, 625--641 (1963).

\bibitem{Aubrey1969}
{D.W. Aubrey}, {G.N. Welding} and {T. Wong}, {J. Appl. Polym. Sci.}, \textbf{13}, 2193--2207 (1969).

\bibitem{Racich1975}
{J.L. Racich} and {J.A. Koutsky}, {J. Appl. Polym. Sci.}, \textbf{19}, 1479--1482 (1975).

\bibitem{Barquins1986}
{M. Barquins}, {B. Khandani} and {D. Maugis}, {C. R. Acad. Sci.
serie II}, \textbf{303}, 1517--1519 (1986).

\bibitem{Maugis1988}
D. Maugis and M. Barquins, Adhesion 12, edited by K. Allen
(Elsevier ASP, London, 1988) pp. 205--222.

\bibitem{Ryschenkow1996}
{G. Ryschenkow} and {H. Arribart}, {J. Adhesion}, \textbf{58}, 143--161 (1996).

\bibitem{Gandur1997}
{M. Gandur}, {M. Kleinke} and {F. Galembeck}, {J. Adhes. Sci. Technol.}, \textbf{11}, 11--28 (1997).

\bibitem{Barquins1997} {M. Barquins} and {M. Ciccotti}, {Int. J. Adhes.}, \textbf{17}, 65--68 (1997).

\bibitem{Ciccotti2004}
{M. Ciccotti}, {B. Giorgini}, {D. Vallet} and {M. Barquins}, {Int. J. Adhes.}, \textbf{24}, 143--151 (2004).

\bibitem{De2004}
{R. De} and {G. Ananthakrishna}, {Phys. Rev. E}, \textbf{70}, 046223 (2004).

\bibitem{Cortet2007} {P.-P. Cortet}, {M. Ciccotti} and {L. Vanel}, {J. Stat. Mech.}, P03005 (2007).

\bibitem{De2008}
{R. De} and {G. Ananthakrishna}, {Eur. Phys. J. B}, \textbf{61}, 475--483 (2008).

\bibitem{Thoroddsen2010}
S.T. Thoroddsen, H.D. Nguyen, K. Takehara, and T.G. Etoth, Phys. Rev. E \textbf{82}, 046107 (2010).

\bibitem{Cortet2013}
{P.-P. Cortet}, {M.-J. Dalbe}, {C. Guerra}, {C. Cohen}, {M. Ciccotti}, {S. Santucci} and {L. Vanel}, {Phys. Rev. E}, \textbf{87}, 022601 (2013).

\bibitem{Marston2014} J.O. Marston, P. W. Riker, and S. T. Thoroddsen, Sci. Rep. \textbf{4}, 4326 (2014).

\bibitem{Dalbe2014}
M.-J. Dalbe, S. Santucci, P.-P. Cortet, and L. Vanel, Soft Matter, \textbf{10}, 132 (2014).

\bibitem{Dalbe2014b}
M.-J. Dalbe, S. Santucci,  L. Vanel and P.-P. Cortet, Soft Matter, \textbf{10}, 9637- 9643 (2014).

\bibitem{video}
We provide in supplementary material a series of videos of typical peeling experiments.  



\bibitem{V_a}
The stick-slip onset velocity $V_{a}$
is slightly larger than in previous
experiments~\cite{Barquins1997,Dalbe2014} where the same adhesive
was peeled directly from its roll. This is
due to the cleaning procedure of the tape backing used as a substrate, which removes its release coating and thus
increases the adhesion. 


\bibitem{Yamazaki2006} {Y. Yamazaki} and {A. Toda}, {Physica D}, \textbf{214}, 120--131 (2006).


\bibitem{Love1944}
{A.E.H Love}, {A Treatise on the Mathematical Theory of Elasticity, Dover Publications}, 1944.

\bibitem{Roman2013}
B. Roman, Int. J. Fract., \textbf{182}, 209--237 (2013).



\bibitem{Weaver1990}
W. Weaver, Jr., S.P. Timoshenko, D.H. Young, Vibration problems in engineering,   Wiley, ISBN: 0471632287, 1990.

\end{thebibliography}
\end{document}